\documentstyle[12pt]{article}
\input{psfig}

\begin{document}

\author{{\bf E. Cuautle}\thanks {e-mail: cuautle@fis.cinvestav.mx}, 
        {\bf G. Herrera} \thanks{ e-mail: gherrera@fis.cinvestav.mx}\\
{Centro de Investigaci\'on y de Estudios Avanzados}\\
{Apdo. Postal 14 740, M\'exico 07000, D.F}\\
\and 
{\bf J. Magnin} \thanks{e-mail: jmagnin@lafex.cbpf.br}\\
{Centro Brasileiro de Pesquisas F\'{\i}sicas} \\
{Rua Dr. Xavier Sigaud 150}\\
{22290-180, Rio de Janeiro, Brazil}}

\title{$D^{\pm}$ and $D^0(\bar{D}^0)$ production asymmetries in $\pi p$
collisions
\thanks{This work was supported by CONACyT (M\'exico) and 
FAPERJ (Brazil).}}

\date{}
\maketitle

\begin{abstract}
We use a two-component model to describe the production 
mechanism of $D$ me\-sons in $\pi p$ collisions. The model combines 
the usual QCD processes plus fragmentation and recombination of charm that
has been produced by nonperturbative QCD mechanisms.
A hard charm component in the pion must be responsible of the particle
anti-particle production asymmetries observed.
\end{abstract}

\newpage

\section{Introduction}

Recent measurements of charm meson production in $\pi p$ collisions 
\cite{e769a,e769b,wa82,e791,wa92} indicate 
that there are important nonperturbative QCD phenomena in the production 
process.\\
The so called "leading effect" or particle anti-particle production 
asymmetry is much bigger than predicted by next-to-leading (NLO),
QCD predictions.\\
In this letter we study the $x_F$ distribution of $D^{\pm}$ and $D^0$
mesons in the framework of a two-component model. The production
of $D$ mesons in the model is assumed to take place {\it via} two different
processes, namely QCD parton fusion with the subsequent fragmentation
of quarks in the final state and conventional recombination of valence
and sea quarks present in the initial state.\\
The asymmetry obtained with the conventional soft charm component
does not reproduce the experimental results.
We show that with the presence of a hard charm component in the pion,
the sum of these two processes gives rise to an enhancement
at large values in the $x_F$ distribution of leading mesons.\\
Unlike the $\Lambda_c$ \cite{nos1,nos2} where the asymmetry is generated
by the
presence of the $ud$ diquark in the initial hadron, the meson
production asymmetry seems to imply the presence of a hard charm component
in the pion.\\
To quantify the difference in the production of leading and non-leading
particles, an asymmetry $A$ is defined 
\begin{equation}
A(x_F)=\frac{\sigma(leading) - \sigma(non leading)}
            {\sigma(leading) + \sigma(non leading)}.
\label{asym}
\end{equation}
In the reaction $\pi(d\bar{u})$ - $proton(uud)$ the $D^0(c\bar{u})$ and 
$D^-(\bar{c}d)$ are leading mesons while $\bar{D}^0$ and $D^+$ are 
non-leading.\\
Several models have been proposed although none fully account for the 
difference in production between leading and non-leading particles. 
The intrinsic charm model \cite{ram-brod} postulates the existence of quantum 
fluctuations in the beam particle that bring about Fock states containing 
$c\bar{c}$ pairs. 
The $c\bar{c}$ quarks having the same velocity as the original
valence quarks are likely to coalesce forming leading particles.
Although the shape of the asymmetry as a function of $x_F$ predicted
by this model is similar to the experimental measurement, the prediction
is to low for the whole $x_F$ interval.\\
In \cite{bednyakov} it is pointed out that the annihilation of the $u$
from the proton and the $\bar{u}$ in the pion would liberate the $d$ of
the pion which can then recombine to form a $D^-$ and certainly not a
$D^+$. The $D^0$ will not be enhanced because it 
cannot be formed with this simple annihilation diagram. The recent measurement
of significant asymmetry in $\Lambda_c$ production in $pp$ collisions
\cite{e769a} however, shows that there must be other production mechanisms
at work that must account for the asymmetry.\\
In two component models \cite{nos1,vogt} the production of charm
mesons by
parton fusion is the same for $D^{\pm}$ and $D^0$. In section 2 the
formalism to obtain the cross section by parton fusion is shown for
charged and neutral charm mesons. 
The recombination of charm as the second component is discussed in
section 3.
Recombination, is different for the charged mesons where
a parton process favors the formation of $D^-$ enhancing the 
$D^{\pm}$ asymmetry. This will be discussed in section 4.

\section{Parton fusion production of charmed mesons}

In the parton fusion mechanism the $D^{\pm}$ $D^0(\bar{D}^0)$  mesons are
produced via the $q\bar{q}(gg) \rightarrow c \bar{c}$ with the 
subsequent fragmentation of the $c(\bar{c})$ quark.
The inclusive $x_F$ distribution of the charm mesons in 
$\pi p$ collisions, is given by 
\cite{vbh-npb} 
\begin{equation}
\frac{d\sigma^{pf} }{dx_F}=\frac{1}{2} \sqrt{s} \int H_{ab}(x_a,x_b,Q^2)
\frac{1}{E} \frac{D_{D/c} \left( z \right)}{z} dz dp_T^2 dy \: ,
\label{sig-qcd}
\end{equation}
where
\begin{eqnarray}
H_{ab}(x_a,x_b,Q^2)& = & 
\Sigma_{a,b} \left( q_a(x_a,Q^2) \bar{q_b}(x_b,Q^2) \right. \nonumber \\
                   &   &
 + \left. \bar{q_a}(x_a,Q^2) q_b(x_b,Q^2) \right) 
\frac{d \hat{\sigma}}{d \hat{t}} \mid_{q\bar{q}} \nonumber \\
                   &   &
 + g_a(x_a,Q^2) g_b(x_b,Q^2) \frac{d \hat{\sigma}}{d \hat{t}}\mid_{gg},
\label{int-qcd}
\end{eqnarray}
and $x_a$ and $x_b$ being the parton momentum fractions, $q(x,Q^2)$ and
$g(x,Q^2)$ the quark and gluon distribution in colliding hadrons, 
$E$ the energy of the produced $c$-quark and 
$D_{D/c} \left( z \right)$ the appropriated fragmentation function. 
In eq. \ref{sig-qcd}, $p_T ^2$ is the squared transverse momentum of 
the produced $c$-quark, $y$ is the rapidity of the $\bar {c}$ quark 
and $z=x_F/x_c$ is the momentum fraction of the charm quark carried by 
the $D$. The sum in eq. \ref{int-qcd} runs over 
$a,b = u,\bar{u},d,\bar{d},s,\bar{s}$.\\
We use the LO results for the elementary 
cross-sections $\frac{d \hat{\sigma}}{d \hat{t}}\mid_{q \bar {q}}$ 
and $\frac{d \hat{\sigma}}{d \hat{t}}\mid_{gg}$:
\begin{equation}
\frac{d \hat{\sigma}}{d \hat{t}}\mid_{q \bar {q}} = \frac{\pi \alpha _{s}^{2}
 \left( Q^2 \right)}{9 \hat{m}_{c}^{4}} \; 
\frac{cosh \left( \Delta y \right) + m_{c}^{2}/ \hat{m}_{c}^{2}}{\left[ 1+
cosh \left( \Delta y \right) \right] ^3}
\label{q-antiq}
\end{equation}
\begin{equation}
\frac{d \hat{\sigma}}{d \hat{t}}\mid_{gg}= \frac{\pi 
\alpha_{s}^{2} \left( Q^2 \right)}{96 \hat{m}_{c}^{4}} \; 
\frac{8 cosh \left( \Delta y \right) -1}{\left[ 1+cosh \left( 
\Delta y \right) \right]^3} \: \left[ cosh \left( \Delta y \right)+
\frac{2m_c^2}{\hat{m}_c^2}+\frac{2m_c^4}{\hat{m}_c^4}\right],
\label{gg}
\end{equation}
where $\Delta y$ is the rapidity gap between the produced $c$ and 
$\bar{c}$ quarks and $\hat{m}_c^2=m_c^2+p_T^2$.  

In order to be consistent with the LO calculation of the elementary 
cross sections, we use the GRV-LO parton distribution functions 
\cite{gr-zpc}, and apply a global factor $K \sim 2-3$ in eq. \ref{sig-qcd} 
to take into account NLO contributions \cite{ramona2}.\\
We take $m_c=1.5 \:GeV$ for the $c$-quark mass and fix the scale of 
the interaction at $Q^2 = 2m_c^2$ \cite{vbh-npb}. 
Following \cite{vogt}, we use a delta fragmentation function 
$D_{D/c}(z) =  \delta(1-z)$, which seems
to describe experimental data better than the Peterson 
fragmentation function.\\

\section{Charmed meson production by recombination}

Sometime ago V. Barger {\it et al.} \cite{halzen} explained the
spectrum enhancement at high $x_F$ in $\Lambda_c$  production 
assuming a hard momentum distribution of charm in the proton.
According with them charm anti-charm pairs which give 
rise to the flavor excitation diagrams (see fig. 1) are not intrinsic but
generated by QCD evolution of the structure functions.\\
In this framework, following the gluon scattering process of fig. 1 the
charm quarks will fragment into charm hadrons. When the $\bar{c}$ is
scattered, the spectator quark could recombine with the $\bar{u}$ valence
quark of the pion to form a $D^0 (c\bar{u})$ or less frequently with an
antiquark from the pion's sea. On the other hand, when the $c$ quark is
scattered, the spectator quark $\bar{c}$ could recombine with $d$ 
valence quark from the pion to form a $D^- (\bar{c} d)$ meson.\\
As pointed out in \cite{halzen} the charm hadrons resulting from the 
scattered charm quark, populate the low $x_F$ region of the cross 
section, while those originating from the spectator quark dominate
at high $x_F$. \\

Here we assume a QCD evolved charm distribution, of
the form proposed by V. Barger {\it et al.} \cite{halzen}
\begin{equation}
xc(x, \langle Q^2 \rangle ) = N x^l (1-x)^k,
\end{equation}
with a normalization $N$ fixed to 
\begin{equation}
\int dx \cdot x c(x) = 0.005
\end{equation}
and $l=k=1$. With this values for $l$ and $k$ one tries to resemble
the distribution of valence quarks. In contrast with the parton fusion
calculation, in which the scale  $Q^2$ of the interaction is fixed at the
vertices of the appropriated Feynman diagrams, in recombination the value
of the parameter $Q^2$ should be used to give adequately the content of
the recombining quarks in the initial hadron. We used $Q^2=4 m_c^2$ and
therefore, the integrated charm quark distribution would take the value
given in eq.7.\\

The production of leading mesons at low $p_T$ was described by 
recombination of quarks long time ago \cite{dh-plb}.\\
In recombination models it is assumed that the outgoing 
hadron is produced in the beam fragmentation region through the 
recombination of the maximum number of valence quarks and the minimum 
number of sea quarks of the incoming hadron.
The invariant inclusive $x_F$ distribution for leading mesons is given by
\begin{equation} 
\frac{2 E}{\sqrt {s}\sigma}\frac{d\sigma^{rec}}{dx_F}=
\int_0^{x_F}\frac{dx_1}{x_1}\frac{dx_2}{x_2}
F_2\left( x_1,x_2\right) 
R_2\left( x_1,x_2,x_F\right) 
\label{rec-cs}
\end{equation}
where $x_1$, $x_2$ are the momentum fractions and
$F_2 \left( x_1,x_2,\right) $ is the two-quark distribution
function of the incident hadron. $R_2 (x_1,x_2,x_F)$ is the 
two-quark recombination function. \\
The two-quark distribution function is parametrized in terms of  
the single quark distributions
\begin{equation}
F_2 \left( x_1,x_2, \right) = 
\beta F_{d,val}\left(x_1\right)F_{c,sea}\left(x_2\right)
\left(1-x_1-x_2\right),
\label{3-quark}
\end{equation}
with $F_{q}\left(x_i\right) = x_iq\left(x_i\right)$.
We use the GRV-LO parametrization for the single quark distributions in
eqs. \ref{3-quark}. The charm contribution however, is parametrized 
with the distribution of eq. 6. It must be noted that since the 
GRV-LO distributions are functions of $x$ and $Q^2$, then our 
$F_2 \left( x_1,x_2 \right)$ also depends on $Q^2$.
The recombination function is given by
\begin{equation}
R_2\left( x_d,x_c\right) =\alpha \frac{x_dx_c}{x_F^2}
\delta \left(x_d+x_c-x_F\right) \: ,   
\label{eq7}
\end{equation}
with $\alpha$ fixed by the condition 
$\int_0^1 dx_F (1/\sigma)d\sigma^{rec}/dx_F = 1$.\\

\section{$D^{\pm}$ and $D^0(\bar{D}^0)$ total production}

The inclusive production cross section of the $D$ is
obtained by adding the contribution of recombination eq. \ref{rec-cs} 
to the QCD processes of eq. \ref{sig-qcd},
\begin{equation}
\frac{d\sigma^{tot}}{dx_F} = \frac{d\sigma^{pf} }{dx_F} + 
\frac{d\sigma^{rec}}{dx_F}. 
\label{sig-tot}
\end{equation}
The resulting inclusive $D$ production cross section 
$d\sigma^{tot}/dx_F$ is used then to construct the asymmetry defined in eq.
\ref{asym}.\\
In the $\pi p$ reaction however, the recombination of a $D^-(d\bar{c})$ gets a
higher probability than the recombination of a $D^0$ $\bar{u}c$ 
\cite{bednyakov}.\\
The process that releases a $d$ quark from the incident $\pi$ to 
form a  $D^-(d\bar{c})$ ( after recombination with a c-quark from the sea )
is also present in the formation of a $D^0$. In this case the released
quark should be $\bar{u}$. In the $D^-$ case however, there is an additional
mechanism that releases the $d$ quark from the pion, namely 
the fusion of a $u$-quark from the proton with the $u$ from the pion.
This is correctly incorporated by the fact that a higher 
contribution from the recombination is needed to describe the $D^-$ asymmetry
while $D^0 (\bar{D}^0$) production and its production asymmetry requires
a small contribution from the recombination process.\\
Fig. 2 shows the model prediction for the $D^-$ and $D^+$ cross section
and experimental results from the WA82 Collaboration \cite{wa82}.
The corresponding production asymmetry is shown in fig. 3 together with 
all the experimental measurements available.
Fig. 4 shows the cross section for $D^0$ and $\bar{D}^0$ mesons
produced in $\pi^-p$ collisions and the experimental points
from the WA92 Experiment \cite{wa92}. 
The corresponding production asymmetry for $D^0$ and $\bar{D}^0$ is shown
in fig. 5. By looking at the fit of the cross section in fig.4 one can see
that the asymmetry obtained for $D^0 (\bar{D}^0)$ could actually change
drastically by slightly changing the curves. More statistics is needed
to improve the confidence of the predicted asymmetry.

\section{Conclusions}

In an earlier work \cite{nos2} the production asymmetry of $\Lambda_c$ was
described using the same recombination scheme used here. The 
presence of a diquark in the initial state, plays an important 
role in $\Lambda_c$ production.
However, it is not possible to describe the
asymmetry in charm mesons using the GRV structure functions for the charm
in the pion. A hard charm component must be present. This fact has been
pointed out before \cite{ram-brod,halzen} but the asymmetry in the
intrinsic charm model \cite{ram-brod} does not seem to fit experimental
results very well. The hard charm component proposed in \cite{halzen}
and a recombination scheme, give a good description of particle
anti-particle production asymmetries for mesons.

\newpage

\section*{Figure Captions}
\begin{itemize}
\item
[Fig. 1:] Heavy flavor excitation diagram. This and similar diagrams are
included in a NLO QCD calculation. However, in oder to reproduce 
particle anti-particle asymetries a hard charm component in the pion 
is needed.
\item
[Fig. 2:] Production cross section for $D^-$ and $D^+$ mesons
in $\pi^-p$ collisions. The dashed line represents the parton fusion 
contribution with a delta fragmentation function. The dotted line gives
the contribution from recombination for $D^-$. The solid line is the sum
of the two contributions.
\item 
[Fig. 3:] Measured production asymmetry for $D^-$ and $D^+$ and
the model incorporating a hard charm component in the pion (solid line).
The dotted line shows the intrinsic charm model prediction.

\item
[Fig. 4:] Production cross section for $D^0$ and $\bar{D}^0$ mesons
produced in $\pi^-p$ collisions. Experimental points \cite{wa92} and 
two component model prediction (solid line). 
\item 
[Fig. 5:] Measured production asymmetry for $D^0$ and $\bar{D}^0$ and
the model incorporating a hard charm component in the pion. The horizontal
line at $A(x_F)=0$ is for reference only,
\end{itemize}

\newpage
\begin{figure}[b]
\psfig{figure=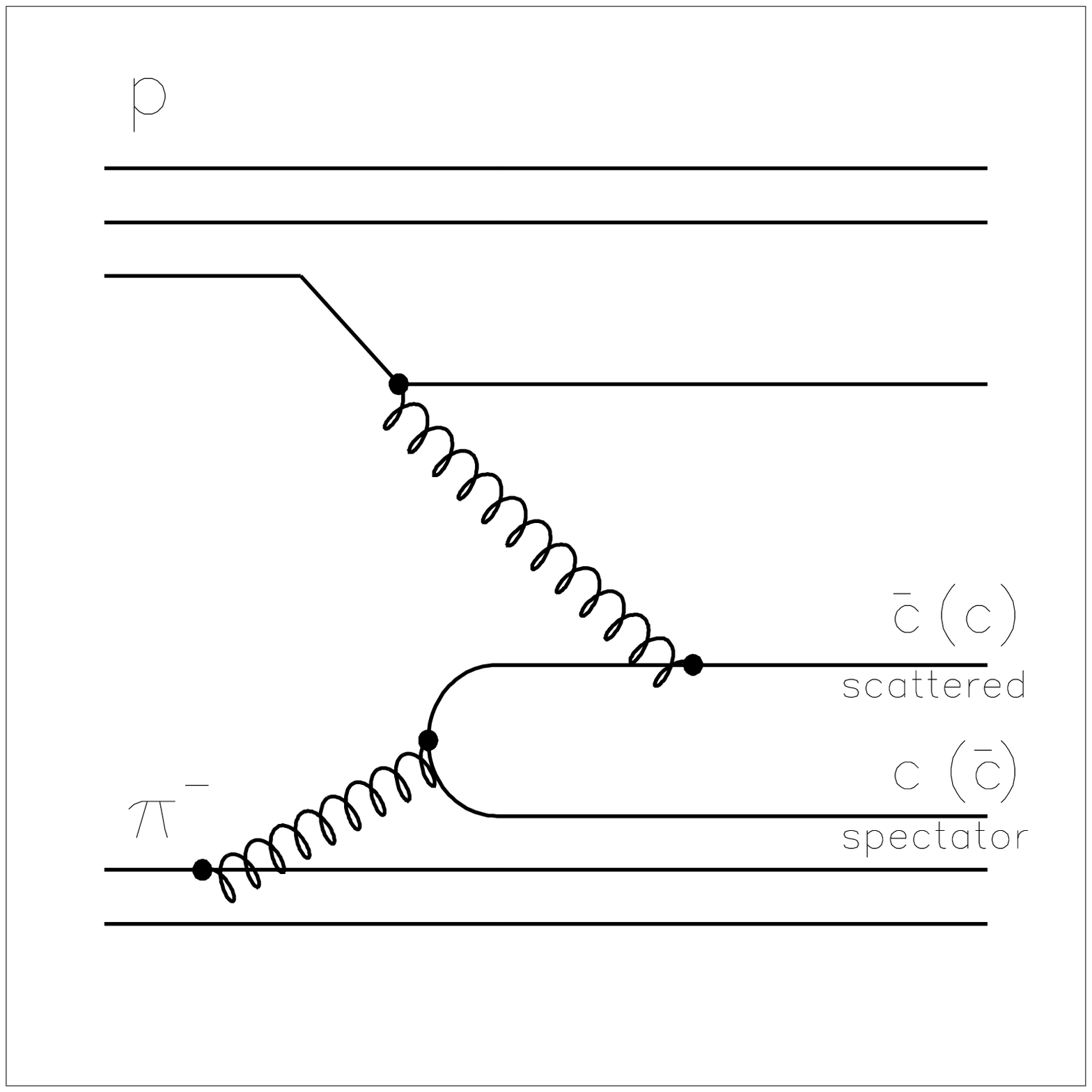,height=6.0in}
\caption{}
\label{dif}
\end{figure}
\begin{figure}[b]
\psfig{figure=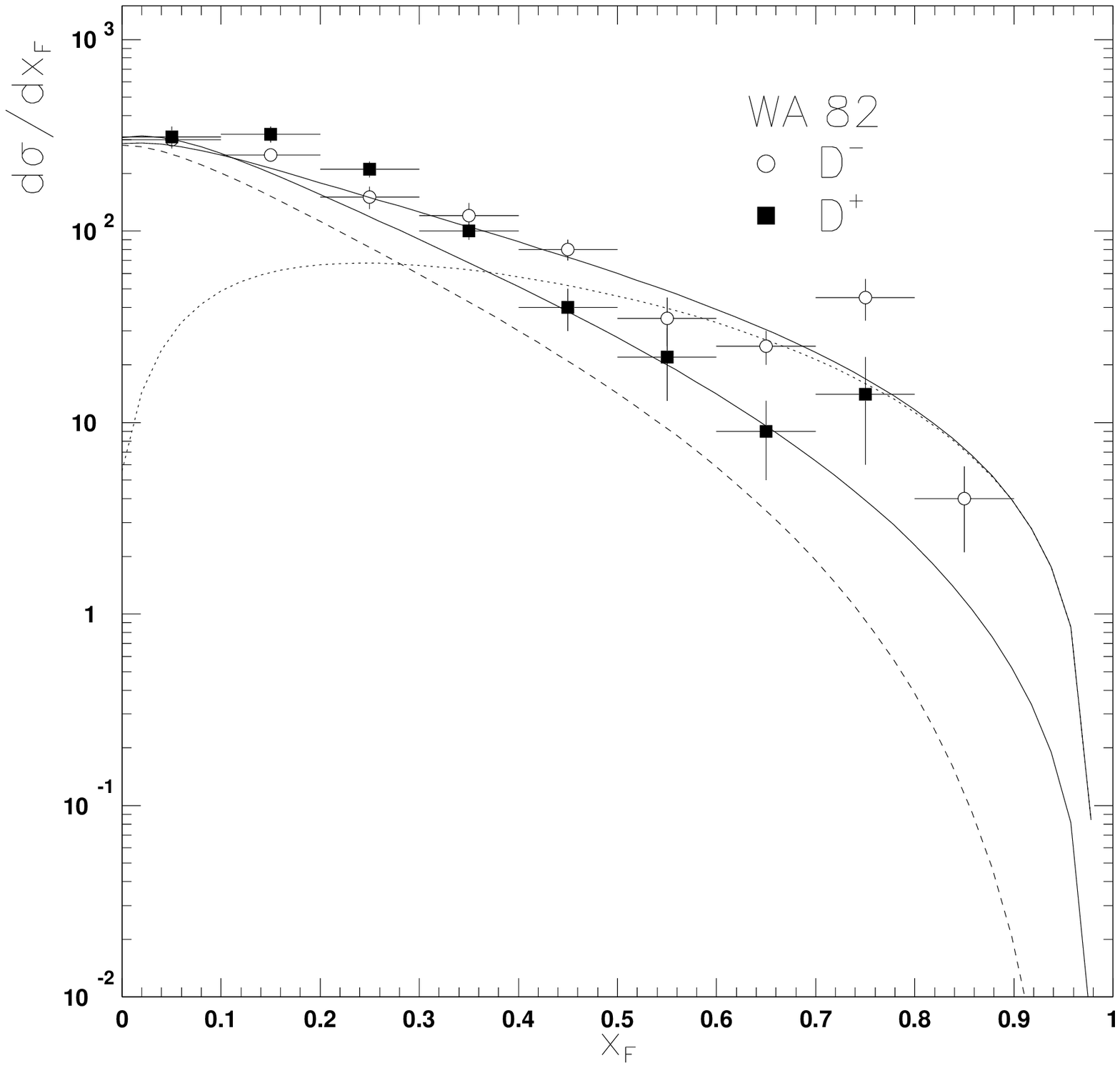,height=6.0in}
\caption{}
\end{figure}
\begin{figure}[b]
\psfig{figure=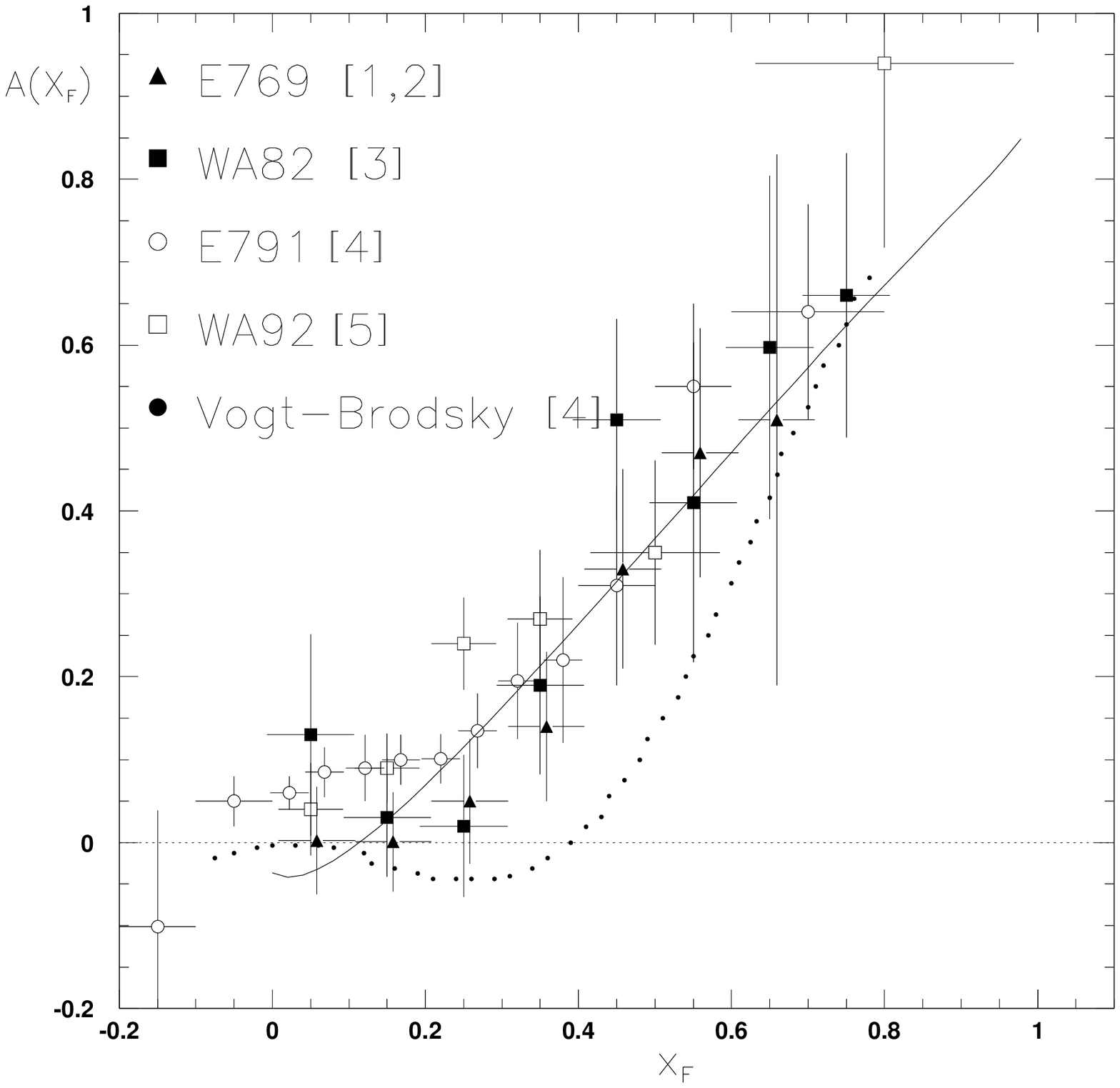,height=6.0in}
\caption{}
\label{asimdd}
\end{figure}
\begin{figure}[b]
\psfig{figure=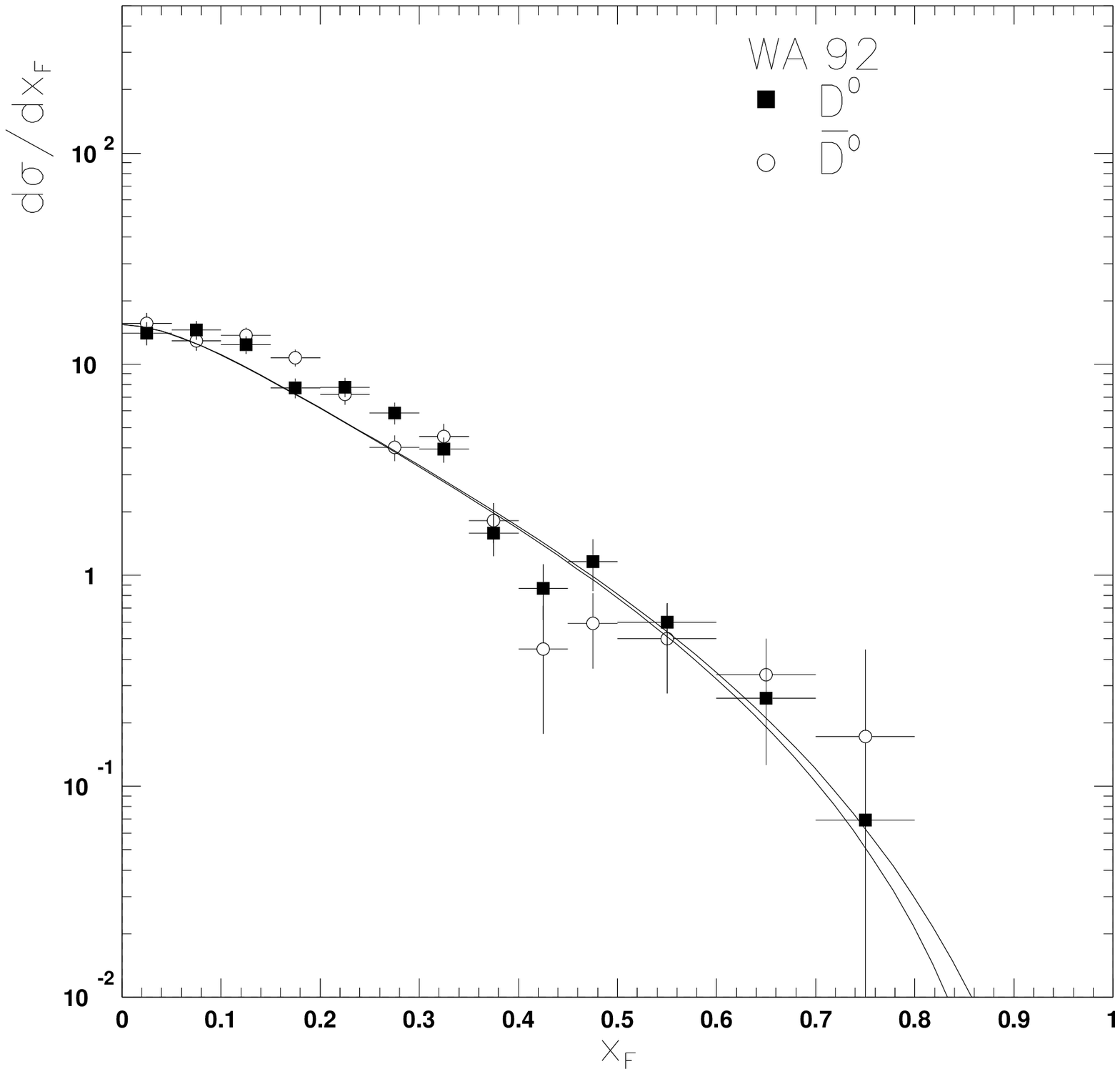,height=6.0in}
\caption{}
\label{prod}
\end{figure}
\begin{figure}[b]
\psfig{figure=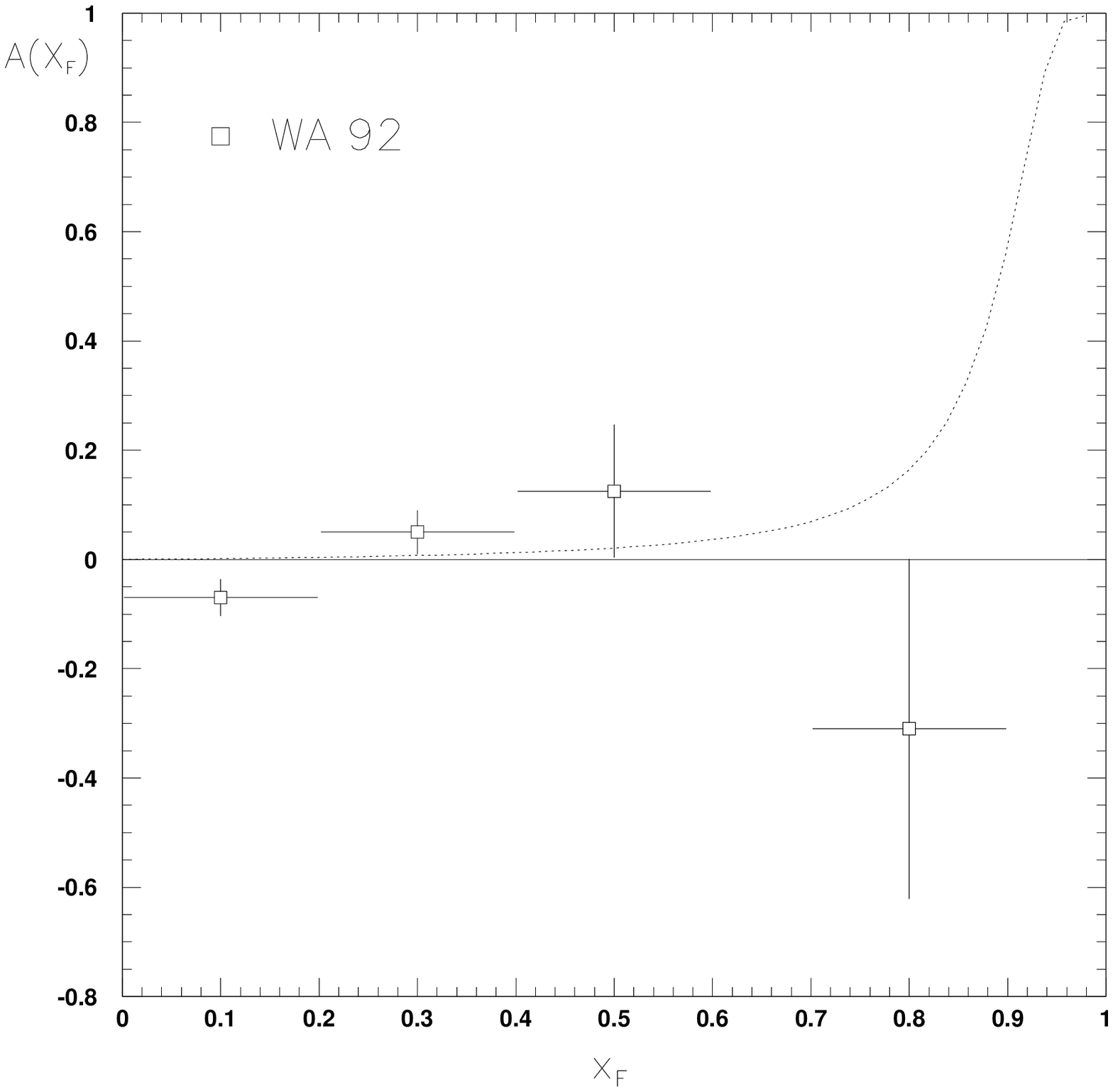,height=6.0in}
\caption{}
\label{asimd0}
\end{figure}
\end{document}